\documentstyle[12pt,epsf]{article}
\title{Coherent Interactions in Nucleus- \\ Nucleus Collisions}
\author{Joakim Nystrand\\[3mm] {\it Div. of Cosmic and Subatomic Physics}\\ {\it Department of Physics, Lund University}\\{\it Lund, Sweden}}
\date{}

\begin{document}

\maketitle

\vspace{-0.6cm}
\begin{center}
{\small\it Presented at the IX$^{th}$ Blois Workshop on Elastic and 
Diffractive Scattering, Pruhonice near Prague,  Czech Republic, 
June 9 -- 15, 2001.}
\end{center}

\begin{abstract}
At the Relativistic Heavy-Ion Collider (RHIC) at Brookhaven
National Laboratory and at the Large Hadron Collider (LHC) at CERN,
particles will be produced in coherent and diffractive nuclear interactions.
In extremely peripheral nuclear collisions (b$>$2R), coherent interactions
occur at very high rates and are dominated by photon-Pomeron
or photon-meson processes. In these reactions, the photon and the
Pomeron/meson from the electromagnetic and nuclear fields couple
coherently to all nucleons. The rates for photonuclear interactions
are roughly two orders of magnitude larger than for two-photon
interactions at comparable center-of-mass energies. 
\end{abstract}

\section{Introduction}

Ultra-relativistic nucleus-nucleus collisions are usually associated with 
copious particle production and complete disintegration of the projectile 
and target nucleus\cite{QM99}. For very high center-of-mass energies, 
a new class of events can be distinguished, however, in which particles 
are produced in very peripheral collisions with no or very little 
disruption of the projectile and target. In these interactions, particles 
are produced through an interaction of the electromagnetic or nuclear fields 
of the ions. If the momentum transfers are small enough ($Q < 1/R$), the 
fields couple coherently to all nucleons, which has the effect of dramatically 
increasing the cross sections. The coherence requirement also gives the events 
a distinct transverse momentum distribution, which can be used to identify 
the events, as will be discussed below.

Coherent electromagnetic interactions can happen for impact parameters of 10's 
or 100's of fermi because of the long range of the electromagnetic field. 
The leptodermous nature of ordinary nuclei, together with their finite spatial 
extent of a few fermi, thus ensures that these interactions can be clearly 
separated from hadronic interactions in impact parameter space. 

The Relativistic Heavy-Ion Collider (RHIC) at Brookhaven National Laboratory 
is the first heavy-ion accelerator energetic enough for significant production
of hadronic final states in coherent nuclear interactions. The first 
collisions at RHIC were achieved in the summer last year, and the STAR 
collaboration has already shown the feasibility of experimentally studying 
these interactions\cite{Janet}. 
The Large Hadron Collider at CERN, when it is operated with heavy-ions, will 
also produce coherent interactions with large cross sections.

In this talk, I will give an overview of two-photon and coherent photonuclear 
interactions in nucleus-nucleus collisions at RHIC and the LHC.

\section{The method of equivalent photons}

The Lorentz contracted electromagnetic fields of a relativistic heavy ion 
can be treated as a stream of equivalent photons. This is the so-called 
Weizs{\"a}cker-Williams method\cite{WW}. In the impact parameter 
representation, the density of photons at a perpendicular distance $b$ 
($b > R$) from the center of the ion is
\begin{equation}
   n(\omega,b) =
   \frac{dN_{\gamma}}{d \omega d^2b} = 
   \frac{\alpha Z^{2}}{\pi^2} \frac{1}{\omega b^2} x^2  K_1^2(x) \; .
\label{WWeq}
\end{equation}
where $\omega$ is the photon energy and $x = b \omega / \gamma$, 
$\alpha$ is the fine structure constant, and 
$K_1(x)$ is the modified Bessel function\cite{Jackson}. Natural units, in 
which $\hbar = c = 1$, are used. Note that the density of photons is 
proportional to $Z^2$. 

The photons from one of the nuclei may interact coherently with the 
electromagnetic or nuclear field of the other nucleus. In the latter case, 
the interactions is of the type photon-meson or photon-Pomeron, whereas 
a purely electromagnetic interaction is referred to as a two-photon
interaction, although higher-order processes are also possible. 

\begin {table} [bth] \begin{center} 
\begin{tabular} { l c c c r c c } \hline 
Accelerator & Ion & A   & Z  & E$_{BEAM}$ [A GeV] & $\gamma$ & Luminosity \\ \hline
RHIC        & Au  & 197 & 79 & 100   \hspace*{1.0cm} & 108.4    & $2 \cdot 10^{26}$ cm$^{-2}$ s$^{-1}$ \\
LHC         & Pb  & 208 & 82 & 2,760 \hspace*{1.0cm} & 2940     & $4 \cdot 10^{26}$ cm$^{-2}$ s$^{-1}$ \\ \hline
\end{tabular}
\label{accelerator}
\caption{Some data for the heaviest beam nuclei at RHIC and LHC.}  
\end{center} \end{table}

The total number of equivalent photons of a given energy is obtained by 
integrating Eq.~\ref{WWeq} over all impact parameters with a suitable 
minimum impact parameter cut-off. For a single nucleus, this cut-off is 
usually given by the nuclear radius, R. The full photon spectrum in a 
heavy-ion interaction is not useable, however, since for impact parameters 
$b<2R$ hadronic interactions dominate and it is generally not possible to 
distinguish electromagnetic interactions. The spectrum for two colliding nuclei can 
then be calculated from\cite{PRC}
\begin{equation}
n(\omega) = 
{dN_\gamma \over d \omega} =  
\int_0^{\infty} 2\pi bdb [1 - P_{Had}(b)]
\int_0^R  {rdr \over \pi R^2} 
\int_0^{2\pi} d\varphi 
\ \ n( \omega , b + r\cos(\varphi) )
\label{PRCdndk}   
\end{equation}
Here, $P_{Had}$ is the probability of having a hadronic interaction at 
impact parameter $b$. $P_{Had}$ is calculated using a Glauber model with 
the total nucleon-nucleon cross section as input. The integrals over $R$ 
and $\varphi$ correspond to an averaging of the flux over the transverse 
surface of the nucleus.  

The total production cross section can calculated as the convolution of 
the photon spectrum with the $\gamma$A photonuclear cross section: 
\begin{equation}
\sigma(A+A \rightarrow A+A+X) = \int n(\omega) \sigma_{\gamma A}(\omega) d\omega .
\end{equation}
The photonuclear cross section for the photon-Pomeron/meson and two-photon 
production mechanisms will be discussed in the next two sections.

\section{Vector meson production}

A photon might interact hadronically by fluctuating into a virtual quark 
anti-quark pair. The $q \overline{q}$-pair prefer to act as a vector meson 
to conserve the spin of the photon. According to the so-called Vector Meson 
Dominance Model\cite{VMD}, the scattering amplitude for photon-hadron 
interactions factorizes into a product of the probability for the fluctuation 
into the vector meson state and a hadronic cross section:
\begin{equation}
\frac{d \sigma}{dt} (\gamma A) = 
\sum_{V} \frac{4 \pi \alpha}{f_V^2} \frac{d \sigma}{dt} (VA) ,
\end{equation}
where $f_v$ is the photon vector meson coupling, and $\sqrt{t}$ is the 
momentum transfer from the target nucleus, summed over the vector meson states. 
Neglecting cross terms (i.e. 
cases where the photon fluctuates into a state V and then scatters off 
the target into a state V'), the hadronic part of the interaction can thus 
be treated as elastic scattering.
For elastic scattering, the photonuclear cross section can be written as a 
product of a constant forward scattering amplitude times an integral over t 
of the nuclear form factor squared, 
\begin{equation}
   \sigma_{\gamma A}(\omega) = \left. \frac{d \sigma}{dt} \right|_{t=0} 
   \int_{t_{min}}^{\infty} | F(t) |^2 dt ,
\label{sigmaga}
\end{equation}
where $\sqrt{t_{min}} = ( M_V^2 / 2 \omega)$ is the minimum momentum transfer 
needed to produce a vector meson with mass $M_V$.
The forward scattering amplitude is related to the total cross 
section through the optical theorem
\begin{equation}
\left. \frac{d \sigma(VA \rightarrow VA)}{dt} \right|_{t=0}  = 
\frac{\sigma^2_{tot}(VA)}{16 \pi} 
\end{equation}

\begin{figure}[t!]
    \epsfxsize=0.9\textwidth
    \centerline{\epsffile{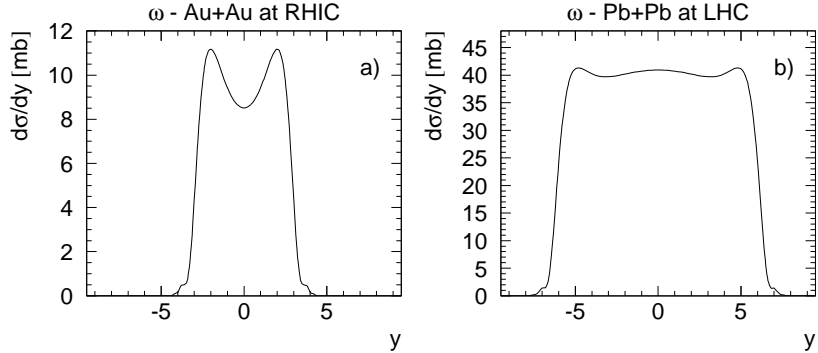}} 
    \label{dndy}
    \caption{Rapidity distributions of $\omega$-mesons at RHIC and LHC.}
\end{figure}

The nuclear forward scattering amplitude can be expressed in terms of the photon-nucleon
cross section, $\sigma(\gamma N)$:
\begin{equation}
\left. \frac{d \sigma(\gamma A \rightarrow VA)}{dt} \right|_{t=0} =
\left( \frac{\sigma_{tot}(VA)}{\sigma_{tot}(VN)} \right) ^2
\left. \frac{d \sigma(\gamma N \rightarrow VN)}{dt} \right|_{t=0}  .
\end{equation}
Measurements of 
coherent photonuclear vector meson production can provide information 
on the total vector meson nucleus cross section, and hence on the 
mean free path of vector mesons in nuclear matter. In ref.~\cite{PRC}, 
data on $\gamma N \rightarrow V N$ from HERA and fixed target 
experiments were used as input to determine 
$\sigma_{tot}(VN)$. This was then used to calculate $\sigma(VA)$ and 
$\sigma(\gamma A)$ from a Glauber model. The corresponding nucleus-nucleus
cross sections were calculated using the photon spectrum of Eq.~\ref{PRCdndk}.
The cross sections for vector
meson production at RHIC and the LHC (Table 2) were found to be very large, 
roughly 10\% of the total inelastic hadronic Au+Au cross section 
at RHIC, rising to 50\% for Pb+Pb collisions at the LHC. 

The vector mesons are produced near mid-rapidity. As an example, the 
rapidity distributions of $\omega$-mesons at RHIC and LHC are shown 
in Fig.~1. 

\begin {table} [hbt] \begin{center} 
\begin{tabular} { l r r r r } \hline 
            & $\rho^0$ & $\omega$ & $\phi$ & $J / \Psi$ \\ \hline
RHIC Au+Au  & 590      & 59       & 39     & 0.29       \\
LHC  Pb+Pb  & 5200     & 490      & 460    & 32         \\ \hline
\end{tabular}
\label{xsect}
\caption{Vector meson production cross section [mb] with heavy ions at 
RHIC and the LHC, from \cite{PRC}.}  
\end{center} \end{table}

\section{Two-photon interactions}

The cross section for meson production through the two-photon channel in 
$\gamma A \rightarrow X A$ interactions is
\begin{equation}
\sigma_{\gamma A}(\omega) = 
8 \alpha Z^2 \frac{\Gamma_{\gamma \gamma}}{M_x^3}
\int \left( \frac{\omega}{Q} \right)^4 | F(Q^2) |^2 \sin^2(\theta) d \Omega ,
\end{equation}
where $\Gamma_{\gamma \gamma}$ and $M_x$ are the two-photon width and 
mass of the meson, $\theta$ is the scattering angle, and $Q$ the momentum
transfer\cite{Budnev}. With a simplified ansatz for the form factor, 
$F(Q^2) = 1$ if $Q < 1/R$ and 0 otherwise, the cross section becomes
\begin{equation}
\sigma_{\gamma A}(\omega) = 16 \pi \alpha Z^2 
\frac{\Gamma_{\gamma \gamma}}{M_x^3} \ln( \frac{2 \omega}{M_x^2 R} )
\end{equation} 
The cross section thus increases logarithmically with the photon energy. 
Using a more realistic form factor
\begin{equation}
F(Q^2) = \frac{3}{(Q R)^3}
\bigg[\sin(Q R) - Q R \cos(Q R)\bigg] \ \ \bigg[{1\over1+a^2Q^2}\bigg] , 
\end{equation}
where a=0.7~fm reflects the nuclear skin thickness, the cross section for two-photon 
production of $\eta$-mesons is compared
with the photonuclear cross section for $\omega$-mesons (Eq.~\ref{sigmaga}) 
in Fig.~2.

\begin{figure}[tbh]
    \epsfxsize=0.7\textwidth
    \centerline{\epsffile{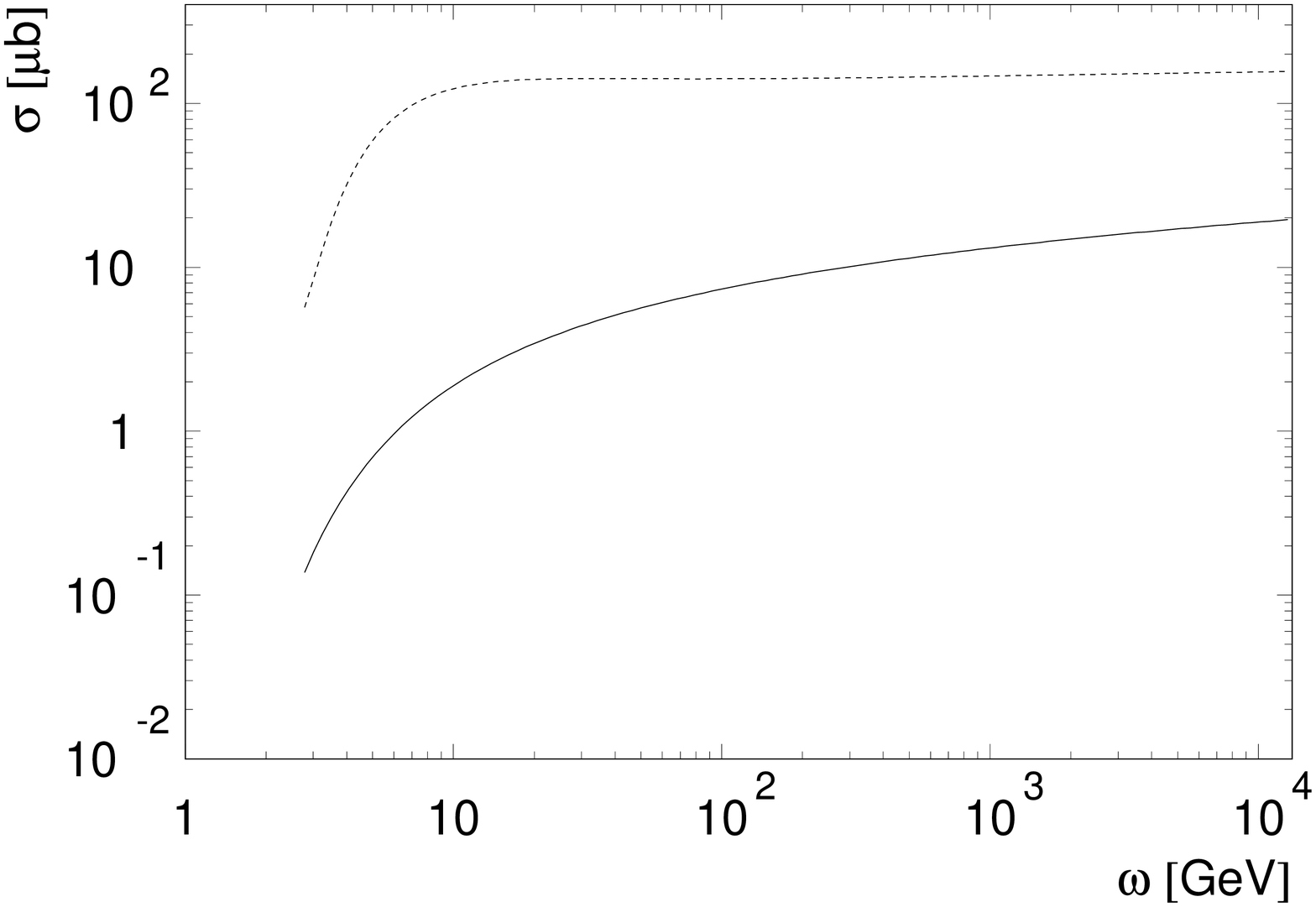}} 
    \label{primakoff}
    \caption{Cross sections for coherent photonuclear production of 
$\omega$-mesons (dotted curve) and two-photon production of 
$\eta$-mesons (solid curve) in $\gamma$+Cu interactions as functions of 
the photon energy, $\omega$, in the nuclear rest frame.}
\end{figure}

The cross section for two-photon production is between one and two orders
of magnitude smaller than that for photonuclear production up to photon 
energies of $10^4$~GeV. 
The energy dependence is very different for the two production mechanisms.
The photonuclear cross section is essentially independent of energy 
above threshold, whereas the two-photon cross section increases 
logarithmically with $\omega$. 

In heavy-ion interactions it is necessary to exclude interactions where
the nuclei overlap to obtain realistic cross sections for 
two-photon processes\cite{BaurCJ}. The differential cross section is given by 
\begin{equation}
   \label{lumeq}
   \frac{ d \sigma}{d W d y} = 
   \frac{W}{2} \int_{b_1>R} \int_{b_2>R} n(\omega_1,b_1)
   n(\omega_2,b_2) \sigma_{\gamma \gamma}  
   \Theta( \mid \vec{b}_1 - \vec{b}_2 \mid - 2R)  d^2 \vec{b}_1  d^2 \vec{b}_2 \; ,
\end{equation}
where $Y$ and $W$ are the rapidity and center of mass energy of produced state.  
The $\Theta$-function removes the contribution from interactions 
with overlap.

\section{Interference and transverse momentum}

It is generally not possible to tag the outgoing nuclei in a coherent 
interaction. The coherence requirement limits the transverse momentum 
transfers to about $\Delta p_T < \hbar c /R$. This means that the angular 
deflection of the ions will be of the order of 
\begin{equation}
   \theta \sim \frac{0.175}{\gamma \cdot A^{4/3}}
\end{equation}
This is a few $\mu$rad at RHIC and a few tenths of a $\mu$rad at LHC. 

Experimental identification of coherent events can instead be achieved 
through the transverse momentum obtained by reconstructing all particles 
emitted in the event\cite{Janet,Lund98}. 

The transverse momentum distribution of virtual photons of energy $\omega$ 
is given by\cite{gammapt}
\begin{equation}
   \frac{dN_{\gamma}}{d k_{\perp}} \propto 
\frac{ | F( (\omega/\gamma)^2 + k_{\perp}^2 ) |^2 }{ ( (\omega/\gamma)^2 + k_{\perp}^2)^2 } k_{\perp}^3 .
\end{equation}
Similarly, the distribution of the nuclear transverse momentum transfer, 
$q_{\perp}$, is given by the form factor\cite{PRL}
\begin{equation}
   \frac{dN_{P}}{d q_{\perp}} \propto  
   | F( t_{min} + q_{\perp}^2 ) |^2 \, q_{\perp} .
\end{equation}

The transverse momentum distribution of the produced state is given by the 
convolution of the transverse momentum distributions of the two sources
\begin{equation}
\frac{dn}{d p_T} = \int f_1 ( \vec{p}_T' ) f_2 ( \vec{p}_T - \vec{p}_T' ) 
d^2 \vec{p}_T' .
\end{equation}

The $p_T$ distribution for states produced in two-photon interactions at
midrapidity is shown in Fig.~3. The shape of the distribution depends on 
the mass of the final state. The corresponding distribution for photonuclear
production of $\phi$ and $J / \Psi$ mesons are shown as the dashed curve in
Fig.~4. The photonuclear $p_T$ distributions are dominated by the nuclear 
form factors and are much less sensitive to the mass of the produced state. 

\begin{figure}[htb]
    \epsfxsize=0.7\textwidth
    \centerline{\epsffile{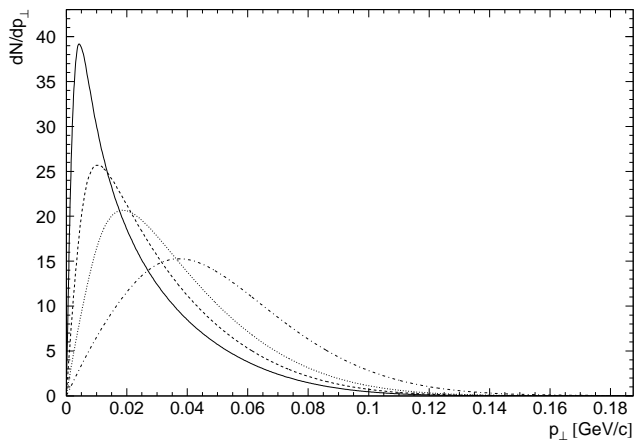}} 
    \label{gammapt}
    \caption{Transverse momentum spectra for two-photon states at midrapidity
for $\gamma \gamma$ center-of-mass energies of 0.2 (solid), 0.5 (dashed), 
1.0 (dotted), and 3.0 (dash-dotted) GeV in Au+Au interactions at RHIC.}
\end{figure}

The situation for photonuclear production is, however, a little bit more 
complicated. Consider the differential cross section for the production
\begin{equation}
\label{intint}
\frac{d \sigma}{dy dp_T} = \int 
\omega_1 \frac{dN}{d \omega_1 d^2b} \, \sigma(\gamma A_2) \, f_{1,2}(p_T) 
\; + \; 
\omega_2 \frac{dN}{d \omega_2 d^2b} \, \sigma(\gamma A_1) \, f_{2,1}(p_T) 
\; d^2b   ,
\end{equation}
where the two terms correspond to production off each of the two nuclei. 

Adding the cross section is only valid as long as $p_T << 1/b$, however, 
since for smaller transverse momenta, the two sources will be 
indistinguishable, and one then has to add the amplitudes\cite{PRL}. At
midrapidity, where the contributions from the two sources are of equal
magnitude, and if one assumes that the outgoing vector mesons can be 
treated as plane waves, the integral over b in Eq.~\ref{intint} becomes
\begin{equation}
   \frac{d \sigma}{dy dp_T} = \int \left| A_1 + A_2 \right|^2 d^2 \vec{b} 
   = 2 \, \int \left| A_1 \right|^2 \, \left( 1 - \cos( \vec{p}_T \cdot \vec{b} ) \right) d^2 \vec{b} . 
\end{equation}
Here, $A_1$ and $A_2$ are the amplitudes for production off nucleus 1 and
2, respectively. This has the effect of modifying the transverse momentum 
spectrum to that shown by the solid curve in Fig.~4. The interference pattern 
is essentially that of a two-source interferometer. 

\begin{figure}[htb]
    \epsfxsize=0.85\textwidth
    \centerline{\epsffile{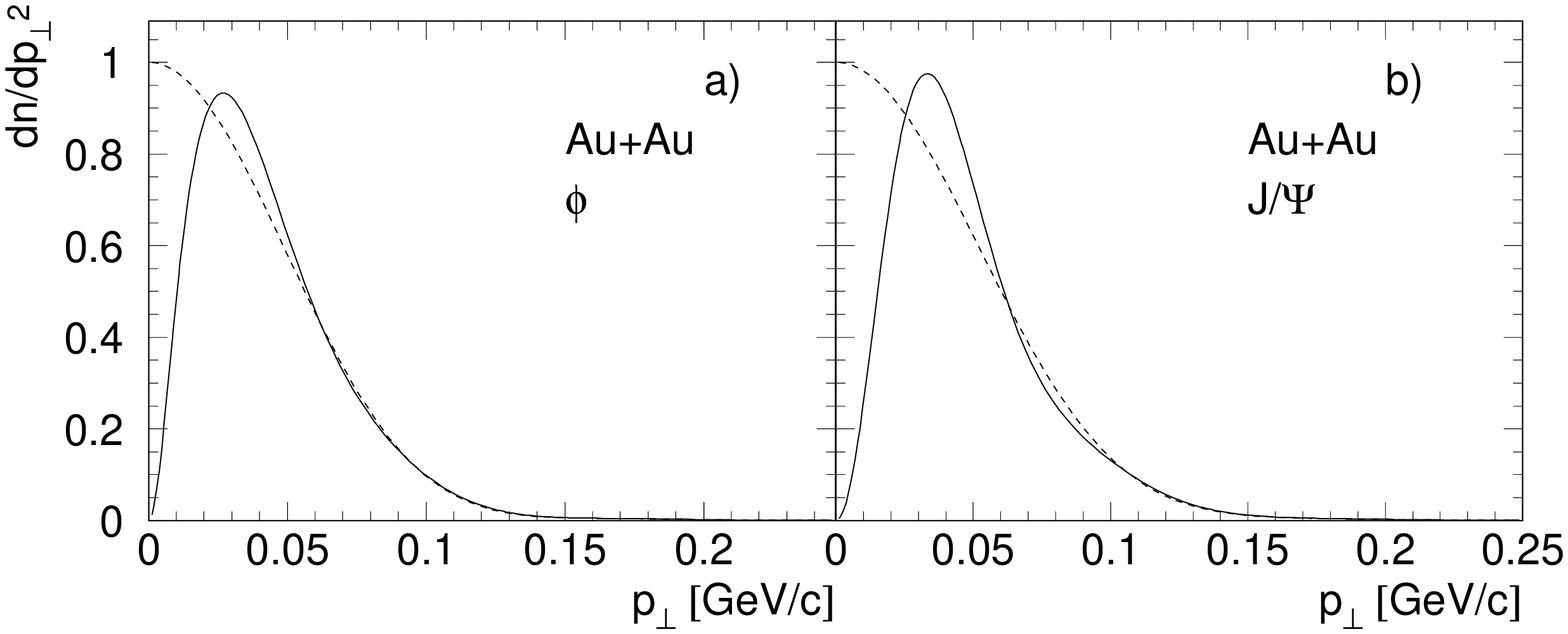}} 
    \label{ptinterf}
    \caption{Transverse momentum spectra with (solid) and without (dashed) 
interference for $\phi$ and $J / \Psi$ production in Au+Au collisions at RHIC.}
\end{figure}

This interference is of particular interest since the $c \tau$ of the 
vector mesons (except for the $J / \Psi$) are generally much shorter 
than the median impact parameters in the interactions. A vector meson
produced at one nucleus  
will thus have decayed before information of its production have 
reached the other nucleus. For interference, the decay 
particles must thus be in an entangled state and retain information about 
their origin long after the decay occurred.

\section{Strong fields and multiple excitations}

Because of the strong fields associated with ultra-relativistic heavy-ions, 
the probabilities for several electromagnetic processes are very large at 
small impact parameters, and calculated, un-unitarized first-order 
probabilities may even exceed 1 \cite{BB}. This is for example the case for 
two-photon production of $e^+ e^-$ pairs. 

Another process with very high interaction probability is mutual Coulomb 
dissociation. The dominating 
process is photonuclear excitation of the target into a Giant Dipole 
Resonance followed by emission of one or more neutrons\cite{BCW}. The 
probability for mutual Coulomb dissociation reaches about 35\% in a 
grazing Au+Au collision at RHIC. 

Coherent vector meson production can occur in 
coincidence with Coulomb excitation of one or both nuclei\cite{Janet}. 
If one assumes that the Coulomb excitation and the vector meson 
production occur independently, the cross section can be calculated from 
\begin{equation}
\label{prob}
\sigma(A+A \rightarrow A^* + A^* + V) = 
\int \, (1 - P_{Had}(\vec{b})) \, P_{Coul}(\vec{b}) \, P_V(\vec{b}) 
\, d^2 \vec{b}
\end{equation}
where $P_{Had}$, $P_{Coul}$, and $P_V$ are the hadronic, Coulomb, and 
vector meson production probabilities, respectively. 

Using the same formalism for vector meson production as in \cite{PRC} 
and the Coulomb reaction probabilities calculated in \cite{BCW},  
Eq.~\ref{prob} gives cross sections of 42~mb for $\rho^0$, 4~mb for $\omega$, 
3~mb for $\phi$, and 0.29~mb for $J / \Psi$ production in Au+Au interactions 
at RHIC. 
The cross sections
are reduced by roughly a factor of 10 compared with no breakup. 

Requiring production in coincidence with nuclear breakup also reduces the 
median impact parameters in the interactions by about a factor of 2. 
This should affect the interference discussed above.

\section{Conclusions}

Electromagnetic interactions will occur with high rates in peripheral 
nucleus-nucleus collisions at RHIC and LHC. The two-photon and coherent
photonuclear production mechanisms have been discussed and compared. 
New phenomena, such as multiple excitations in strong fields and interference, 
not accessible in similar reaction channels in $e^+ e^-$ and $e$A interactions, 
can be studied with heavy ions.

\newpage

\vspace{0.5cm}
\noindent
{\Large\bf Acknowledgements}

\vspace{0.2cm}
\noindent
I would like to acknowledge Spencer Klein, LBNL, Berkeley, my collaborator 
in the studies of vector mesons and interference. I would like to thank 
Tony Baltz and Sebastian White, BNL, Brookhaven, for providing the 
Coulomb interaction probabilities from their paper\cite{BCW} and for useful 
discussion. This project was supported by the Swedish Research Council (VR).

\end{document}